Original Paper

# Extraction of diverse gene groups with individual relationship from gene co-expression networks

Iori Azuma[1, †], Tadahaya Mizuno[1, †, *] and Hiroyuki Kusuhara[1, *]

[1]Department of Pharmaceutical Sciences, The University of Tokyo, Bunkyo, Tokyo, Japan
* To whom correspondence should be addressed. Tel: +81 3 5841 4771; Fax: +81 3 5841 4766; Email: tadahaya@mol.f.u-tokyo.ac.jp. Correspondence may also be addressed to kusuhara@mol.f.u-tokyo.ac.jp. [†]These authors contributed equally to this work.

**Abstract**
**Motivation:** Modules in gene co-expression networks (GCN) can be regarded as gene groups with individual relationships. No studies have optimized module detection methods to extract diverse gene groups from GCN, especially for data from clinical specimens.
**Results:** Here, we optimized the flow from transcriptome data to gene modules, aiming to cover diverse gene–gene relationships. We found the prediction accuracy of relationships in benchmark networks of non-mammalian was not always suitable for evaluating gene–gene relationships of human and employed network-based metrics. We also proposed a module detection method involving a combination of graphical embedding and recursive partitioning, and confirmed its stable and high performance in biological plausibility of gene groupings. Analysis of differentially expressed genes of several reported cancers using the extracted modules successfully added relational information consistent with previous reports, confirming the usefulness of our framework.
**Availability:** The proposed module detection method is available at https://github.com/mizuno-group.
**Contact:** tadahaya@mol.f.u-tokyo.ac.jp

## 1 Introduction

The transcriptome data of a specimen well reflect its properties because they comprise a complete and low-level representation of biological information, considering the central dogma. Given their nature, transcriptome profiles contribute to a wide variety of fields including clinical situations (Li et al., 2018; Morita et al., 2020). A gene group—its signature—is a widely employed form of transcriptome profiles. Differentially expressed genes (DEGs) are often employed, probably because they are simple and easy to prepare: thus, several samples with controls are sufficient to calculate DEGs (Hou et al., 2010). Gene ontology (GO) and pathway analysis are well-established approaches after the acquisition of a gene signature to estimate what the gene group might mean biologically (The Gene Ontology Consortium, 2019; Tian et al., 2005). However, these analyses identify the labels of biological systems and do not provide details on individual relationships between the genes in each group. Therefore, it is difficult to grasp the molecular machinery behind a gene signature by GO and pathway analyses unless the hit terms have been established well. Clinical specimens are precious and limited compared with experimentally prepared specimens, which motivated us to maximize information extracted from a gene signature with similar frameworks by leveraging existing knowledge such as GO analysis.

After epochal studies in the early 2000s, the notion that molecules in an organism comprise a complex network is accepted widely (Barabási and Albert, 1999; Watts and Strogatz, 1998). Genes form a complex network as well, such as gene co-expression network (GCN), and mathematical approaches to network analysis have been shown to contribute to the life sciences (Maertens et al., 2018; Batada et al., 2006). Complex networks contain modules, defined as the division of network nodes into groups where the connections are dense (Saelens et al., 2018; Newman and Girvan, 2004). Modules from GCNs can be regarded as gene groups with relational information and can be useful in the analysis of a gene group of interest in detail, by matching the modules with the gene group, as in GO and pathway analysis. Then, the important point is how to detect reliable and diverse modules from GCNs for this purpose.

The process from transcriptome data to modules is divided into three steps: relationship inference, network construction, and module detection (Figure 1). The former two steps are often integrated into GCN construction algorithms and are evaluated by the same metric: prediction accuracy of relationships or edges. A global competition called the DREAM challenge boosted this area, and winning methodologies such as GENIE3 (Huynh-Thu et al., 2010) have contributed to varying fields (Hossain et al., 2021; Cao et al., 2021). However, the competition employed relatively clean data derived from simulation or non-mammalian models, and extrapolation to complex human networks should be done carefully considering the noisy nature of data from clinical specimens. The methodologies for the last step—detection of modules—are evaluated with biological plausibility of groupings by consistency with existing biological knowledge of gene groups such as in GO databases. However, the instability of hit ratio against the hit number and overlap between the members of the registered groups in existing databases are often overlooked (Song and Zhang, 2015).

Overall, the process from transcriptome data to modules should be evaluated carefully with an awareness of the three sequential steps (relationship inference, network construction, and module detection) and the two different views of biological plausibility (individual relationships and groupings), especially when dealing with human clinical data. However, to the best of our knowledge, there have been no studies that evaluated these points collectively to obtain modules from transcriptome data. Here, we tackled this issue and optimized the above flow, focusing on data from clinical specimens and the diversity of the outcome modules to cover diverse gene–gene relationships in human.

## 2 Methods

### 2.1 Relationship inference methods

Among many algorithms for inferring relationships, we selected the algorithms that would cover the most gene–gene relationships. Details are given in the Supplementary Methods.

**2.1.1 Tree-based method.** Gene Network Inference with Ensemble of Trees (GENIE3) (Huynh-Thu et al., 2010) is a representative algorithm using Random Forest and won the DREAM4 and DREAM5 challenges (Marbach et al., 2010). In this study, we set all the genes as candidate regulators and their target genes.

**2.1.2 Linear regression method.** Elastic Net is formulated as:

$$argmin\{\frac{1}{N}\sum_{i=1}^{N}(y_i - \beta_0 - \sum_{j=1}^{p}\beta_j X_{ij})^2 + \lambda \sum_{j=1}^{p}\left[\frac{1}{2}(1-\alpha)\beta_j^2 + \alpha|\beta_j|\right]\}$$

We set $\lambda = 1$ and $\alpha = 0.005$ so that we could extract sufficient gene–gene relationships for network inference. All genes were used as candidate explanatory variables. Regression analysis was performed for each gene, and the absolute values of the coefficients of the explanatory variable genes were defined as the weight of each relationship.

**2.1.3 Correlation-based method.** Because transcriptome data tend to fluctuate, we focused on the FDR-corrected correlation coefficient, which reduces the impact of false positives with permuting across samples, as a counterpart for the above machine learning methods (Song and Zhang, 2015). The FDR threshold was set to <0.05 in this study.

### 2.2 Network construction (PMFG)

We aimed to construct a planar maximally filtered graph (PMFG) and to extract gene groups as modules (Tumminello et al., 2005). A framework applying PMFG to gene expression was developed in Multiscale Embedded Gene Co-expression Network Analysis (MEGENA) (Song and Zhang, 2015) and we followed this method.

### 2.3 Evaluation metrics of networks

After transforming each network inference to a PMFG, we assessed its complex network properties and gold standard reflection. Details are given in Supplementary Methods.

Complex network properties. Because complex networks are scale-free and small-world, we used them as indicators to assess the quality of the network (Barabási and Albert, 1999; Watts and Strogatz, 1998).

**2.3.1 Shortest path distance (SPD) comparisons.** We evaluated the significant difference between the target and nontarget genes by calculating their SPD from the TF. First, the mean SPD from the TF was calculated for its gold standard target gene. Then, the nontarget genes were randomly selected 10,000 times to be the same number as the target genes and their mean SPD was calculated each time, from which bootstrap $P$ values were calculated.

**2.3.2 TF proximal overlap.** We focused on each TF and investigated its proximal genes. Genes within the second- or third-degree connections from the TF were regarded as its target genes. The significance of the overlap between these proximal genes and gold standard target genes was calculated using FET.

### 2.4 Module detection

In addition to existing and well-known methods for module detection, we tested a new algorithm for GCNs utilizing node2vec, a graph embedding method using the skip-gram algorithm (Grover and Leskovec, 2016). Each GCN was converted into a vectorized matrix using node2vec, keeping the information on network structure, and then unsupervised clustering with recursive partitioning based on Newman's modularity (Q) was conducted as follows:

1. The GCN was transformed to d-dimensional feature vectors about n genes with node2vec.
2. The k-means clustering was repeated while increasing the k-split number one by one and Q values were calculated in each split until no better value was returned for the indicated consecutive times.
3. For the clusters defined above, we performed hierarchical clustering recursively. The minimum cluster size threshold and the degree of compactness were considered to determine whether the next split was allowed. After splitting with the optimal k-split, we checked for each new cluster as to whether it met the minimum cluster size threshold and calculated the compactness degree as follows: $v_l = \frac{SPD \in V_l}{log(|V_l|)}$. The split was rejected based on the percentage of new clusters that met the minimum cluster size and compactness compared with the parent cluster.

Details for the existing methods and other developmental processes are described in Supplementary Methods. The computer code produced in this study is available in Supplementary Code and in the following database: GitHub (https://github.com/mizuno-group/n2v_rc).

### 2.5 Evaluation of module detection

A curated GO dataset was obtained from Enrichr and employed in GO analysis (Kuleshov et al., 2016). By varying the adjusted $P$ value threshold, we calculated the numbers and ratios of modules that showed significant overlap with GO. In addition, we focused on the GO terms that corresponded significantly with the detected modules and evaluated the coverage of their diversity upstream in the GO hierarchy (termed the diversity index). The hierarchical structure of GO was downloaded from the GO consortium (The Gene Ontology Consortium, 2019) and the Core ontology dataset was employed.

### 2.6 Analysis of DEGs by predefined gene groups with relationships

To evaluate the usefulness of the predefined gene groups generated by the proposed framework, we mapped DEGs onto the generated gene groups and visualized them. DEGs derived from the specimens with a similar physiological background to TCGA data were collected and significant enrichment with the generated gene groups was checked using FET. DEGs and the gene groups significantly enriched with the DEGs were visualized as networks and their relationships were investigated in the literature. Details for DEGs analysis are described in Supplementary Methods.

## 3 Results

### 3.1 Overview

Of the three steps from transcriptome data to modules (Figure 1), a planar maximally filtered graph was selected for network construction because it does not require arbitrary hyperparameters (Tumminello et al., 2005). We started by refining the method to infer the relationships between genes and tested three different types of inference that would cover a large portion of gene–gene relationships. The first one is GENIE3 (Huynh-Thu et al., 2010) as a tree-based machine learning method (Tree-based) because it achieved excellent results in the DREAM5 competition (Marbach et al., 2012). We considered that other machine learning methods could be useful in relationship inference as well and selected Elastic Net as another approach (Linear Regression) thanks to its robustness against multicollinearity and having a group effect. In addition, we selected an approach that considers the effect of measurement errors in "omics" data by correcting it with the false discovery rate (FDR) as a competitor to the above two types of machine learning (Correlation-based) (Song and Zhang, 2015). The performance of an inference method is generally evaluated with prediction accuracy of given relationships between transcription factors (TFs) and their target genes. We utilized two datasets of TFs and their target genes: a well-known benchmark from the DREAM5 challenge (Marbach et al., 2012) and the relationships extracted from ChIP-Atlas (Oki et al., 2018). Surprisingly, the performance of the three inference methods did not match between the DREAM5 and ChIP-Atlas datasets, which imply that widely used evaluations based on prediction accuracy are not always suitable for optimization of the methods to infer the relationships between genes (Supplementary Figures S1-S4; Supplementary Tables S1 and S2).

### 3.2 Comparison of inference methods after network construction

Considering our goal, desirable methods of relationship inference can detect gene relationships that contribute to constructing a GCN with high biological plausibility and diversity (Figure 1). Therefore, we evaluated the three inference methods based on the properties of networks derived from the relationship generated by each method. The average shortest path distance (SPD) and the clustering coefficient of GCNs derived from the Correlation-based method were smaller and higher, respectively, than those of GCNs from the other methods. This agreed with the scale-free property as shown in Supplementary Figure S5 and indicates that the Correlation-based GCN well satisfied complex network properties (Figure 2A and Table 1) (Humphries and Gurney, 2008). Next, we assessed whether the target genes of a given TF were concentrated near to it using the gold standard defined with ChIP-Atlas. The average SPD between a TF and its target genes was compared with that between the TF and randomly selected nontarget genes as depicted in Figure 2B. Clear differences were observed in the distribution of degrees for TFs such as FOXM1 and E2F1, which are important genes in GBMs (Figure 2C) (Zhang et al., 2012; Xu et al., 2018). We counted the numbers of TFs whose average SPDs of their target genes were statistically smaller than the nontarget one by calculating the bootstrap $P$ value. As shown in Table 2, the highest number of TFs was given by the Correlation-based method, followed by the Tree-based and Linear Regression methods. In addition, the enrichment of target genes of a TF in the proximity of the TF up to the second and third order was tested using Fisher's Exact Test (FET). Consistent with the results of SPD comparison, GCNs derived from the Correlation-based method had the best correspondence with the gold standard (Supplementary Table S3). These results indicate that the Correlation-based method provided the most desirable inferred relationship for constructing a GCN covering a large portion of gene–gene relationships from the data of clinical specimens.

### 3.3 Comparison of module detection methods

In analyses of biological networks, correspondence with predefined gene groups based on domain knowledge is evaluated for comparing module detection algorithms. The ratio of modules that show significant correspondence is often utilized, whereas the results are unstable depending on the number of modules extracted. For this reason, the number of detected modules that show significant correspondence is also used frequently. However, the independence of each gene group is not always guaranteed in the existing databases of gene groups, and the number of significantly correspondent modules could be overestimated. In fact, when we tested for overlap between each of the registered gene groups in several well-known databases, many of the groups exhibited significant overlap with other groups in the same database (Supplementary Figures S6A and B). To adjust for overlap and cover a large portion of gene–gene relationships, we devised a diversity index that utilizes the structural information of GO definitions, and employed all of the above metrics for comparison of module detection methods (Supplementary Figure S6C).

Module detection methods based on graph embedding such as node2vec have been attracting attention (Grover and Leskovec, 2016). Because there are no examples of module detection based on node2vec for GCNs, we have developed a new method using an algorithm (Figure 3A). It is necessary to perform clustering on the embedded vectors, where determining the number of clusters is one of the outstanding issues. Therefore, we have introduced a method to calculate k-means recursively based on Newman's modularity Q (Newman, 2006), an important measure of network division. In these evaluation metrics, the recursive clustering method outperformed the existing three clustering methods, Affinity Propagation (Frey and Dueck, 2007), DBSCAN (Ester M, Kriegel H-P, Sander J, 1996), and X-means (Dau Pelleg, 2000), which do not require specifying the number of clusters in advance (Supplementary Figure S7, Supplementary Note). Then, we compared the proposed node2vec-recursive clustering algorithm with other well-known algorithms for module detection such as MEGENA (Song and Zhang, 2015), Infomap (Rosvall and Bergstrom, 2008), Spinglass (Reichardt and Bornholdt, 2006), and Spectral Clustering (Andrew Y. Ng, Michael I. Jordan, 2001). As shown in Figure 3B, node2vec-recursive clustering exhibited the most consistent results (Figure 3B, Table 3, Supplementary Figure S8 and Supplementary Table S4). In this way, we have established a framework for detecting diverse gene groups with relational information from the transcriptome data of clinical specimens.

### 3.4 Utilization of diverse gene groups with an individual relationship

DEGs provide information on specimens and can be obtained from relatively few data. In general, DEGs are subjected to GO or pathway analysis to deduce related functions as a system. However, the information provided by these analyses mainly includes the labels of systems such as GO terms, and details of the relationship between the members of DEGs are not provided (Supplementary Figure S9). Mapping DEGs to a set of genes with relational information extracted from a GCN prepared beforehand is expected to help us to understand the relationships between individual genes in the DEGs (Supplementary Figure S10). A previous study (Hou et al., 2010) obtained DEGs of LUAD against normal lung samples by transcriptome analysis. Note that the datasets used to define these DEGs are independent of the datasets used to construct the LUAD GCN in Figure 2. We tested the correspondence between the DEGs and 448 gene groups extracted with our framework using FET, and the DEGs were found to be associated with 15 gene groups (Figure 4A). Focusing on the module that showed the most significant overlap (FDR-corrected $P$ value: $1.3 \times 10^{-27}$), we detected a relationship between 47 members of DEGs as shown in Figure 4B. Notably, the central gene in the module, MFAP4, was not referred to in the previous report (Hou et al., 2010) but has been reported to be a prognostic biomarker of lung cancer (Feng et al., 2020; Yang et al., 2019). In addition, AGER, highly central in the 4th significant module (FDR-corrected $P$ value $6.8 \times 10^{-16}$), coincided with five genes reported to be important by the reference study with a different approach (Figure 4C). Similar results about DEGs and gene groups with relational information generated by the established platform were obtained for HHC and GBM (Supplementary Figure S11). Although these results are limited examples, they support the idea that predefined gene groups with relational information, such as the outputs of our framework, make it possible to infer the relationships between important genes of specimens, even when network analysis is difficult because of the small number of samples.

## 4. Discussion

Today, "omics" approaches are widespread and transcriptome profiles are frequently utilized in various fields of life science. Gene signatures such as DEGs can be obtained easily from clinical specimens and are often reported to be useful in the stratification of disease subtypes and prognosis (Bidkhori et al., 2018; Zuo et al., 2019). However, the reproducibility of these studies is questioned frequently, which is mainly due to the vulnerability of "omics" to unrecognized clinical confounding factors (Song et al., 2020). Therefore, it is important to grasp the machinery behind a gene signature, and its consistency with the application should be checked. GO and pathway analyses are used widely to elucidate such machinery although these analyses return only the names of biological systems and do not provide detailed individual relationships between the genes involved. The proposed framework in this study provides predefined modules derived from GCNs, which could contribute to uncovering the machinery behind a gene signature.

Regarding machine learning algorithms, we selected Random Forest and Elastic net as representative machine learning approaches because these were robust against multicollinearity and have a group effect so were appropriate for the construction of comprehensive gene–gene relationships whereas other algorithms may be utilized for other purposes. Our results point out the pitfalls in the evaluation and optimization of the algorithms for constructing GCNs from the transcriptome data of clinical specimens by using clean sources such as the DREAM5 dataset.

How to evaluate the performance of algorithms for constructing GCNs is often a matter of debate. The predictive accuracy of gold standards is one of the major evaluation metrics in the field of network analysis but the reliability of such standards for biological networks should be considered carefully as described in Supplementary Note. Therefore, we scored the other network-based metrics as well and confirmed that all the different kinds of evaluation metrics—both dependent on and independent of the gold standard—agreed with each other in the comparison of relationship inference methods (Figures 3 and 4). Not only the prediction accuracy of relationships and edges but also the other metrics utilizing network properties are options for evaluating the performance of algorithms for constructing GCNs: human ones in particular.

Evaluation of the performance of the algorithms for module detection should be done carefully. In the field of network analysis, how correctly the predefined groupings are reconstructed is a widely utilized metric. In contrast, because it is quite difficult to define biologically "correct" groupings of genes, the biological plausibility of groupings based on consistency with existing biological knowledge of gene groups is utilized as an alternative evaluation regarding GCNs (Song and Zhang, 2015). We wish to note the overlap between the members of registered groups in existing databases, which can bias existing metrics: the hit number and ratio (Supplementary Figure S6). Thus, we have introduced a new index to overcome this overlap bias. This index—diversity—utilizes the tree structure of the GO database provided by the GO consortium. It scores how diversely the given gene groups cover those defined in the GO database, keeping the same detection sensitivity of group matching (Ashburner et al., 2000). Note that the diversity index might not be appropriate for general use because it was devised to extract diverse gene groups with individual relationships for the present study. We have also introduced a new module detection method utilizing the combination of graph embedding and recursive partitioning of networks using modularity (Grover and Leskovec, 2016). The latter approach for GCNs was devised by Song et al. in a sophisticated way whereas the application of graph embedding by node2vec for module detection from GCNs has not been reported. Graph embedding by node2vec transforms GCNs into vectors of nodes while allowing for adjustment of the depths and widths to be considered between nodes. It expands the clustering options in subsequent recursive partitioning, which are not possible with the original approach using k-medoids clustering. In this study, we demonstrated the usefulness of this flexible approach in extracting diverse gene groups from GCNs, and its applications for other purposes remain to be explored.

### Acknowledgements

We thank the investigators who obtained transcriptome profile data in the TCGA project (Weinstein et al., 2013).

### Funding

This study was supported by a Grant-in-Aid for Scientific Research (C) [21K06663] from the Japan Society for the Promotion of Science and a Grant-in-Aid from the Takeda Science Foundation.

## Conflict of Interest

None declared.

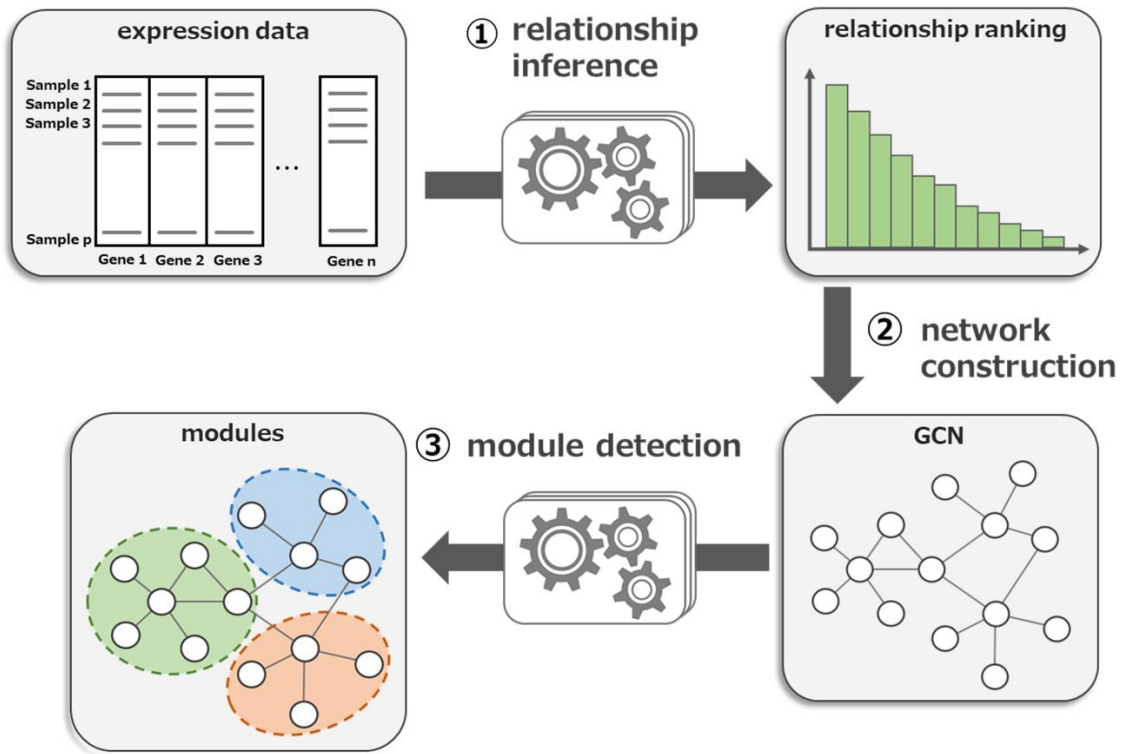

**Fig. 1.** A framework for inferring pairwise relationships from gene expression data, constructing gene co-expression networks, and extracting a set of genes with relational information.

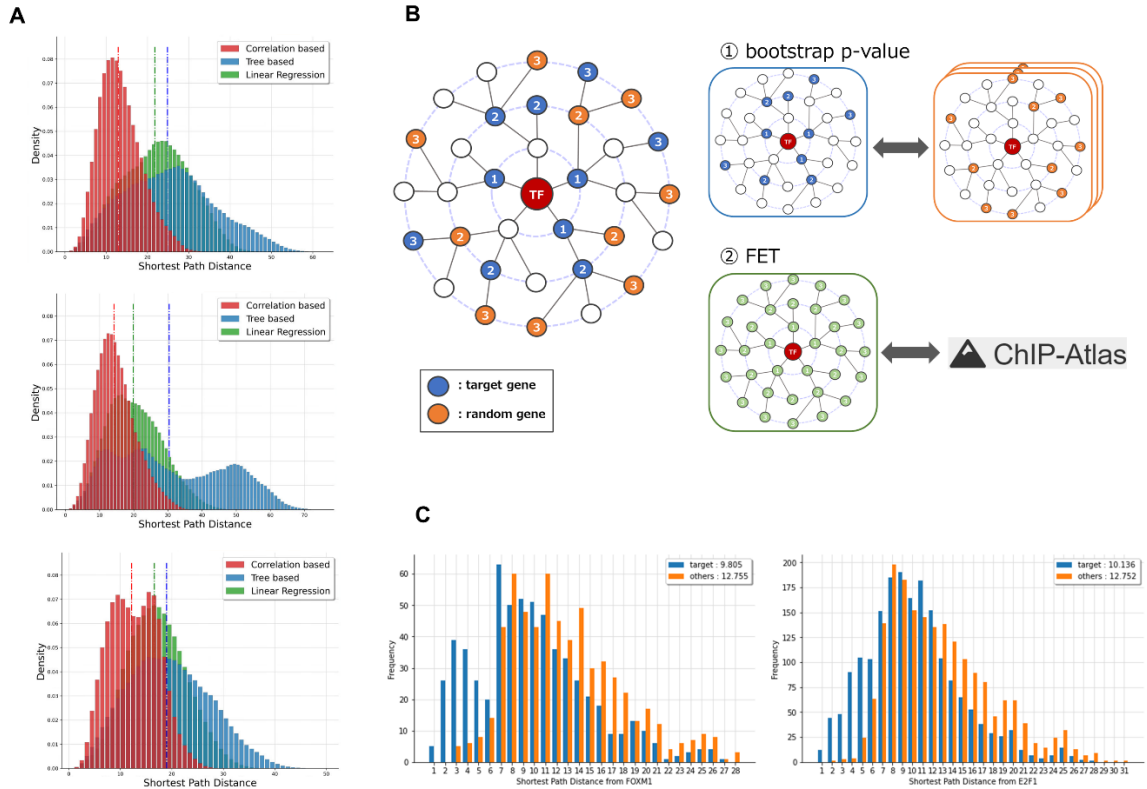

**Fig. 2. Comparison of inference methods after network construction.** (A) Histogram of the shortest path distance (SPD) of each network constructed with Correlation-based (red), Tree-based (blue), and Linear Regression (green) methods. 1,000,000 gene pairs were selected randomly, and the SPD is depicted. The dashed line shows the average SPD in each network. (B) A method to compare whether the target genes (orange) are concentrated in the proximity of their transcription factor (TF, red) compared with randomly selected nontarget genes (orange). The enrichment of target genes in the proximity of the TF was evaluated by two methods; comparing the average SPD with bootstrap *P* values, and performing FET for testing the overlap with the gold standard derived from ChIP-Atlas. (C) Enrichment of target genes in the proximity of FOXM1 (left) and E2F1 (right), both of which are important in GBMs. The target genes were clearly concentrated in the vicinity of their TFs.

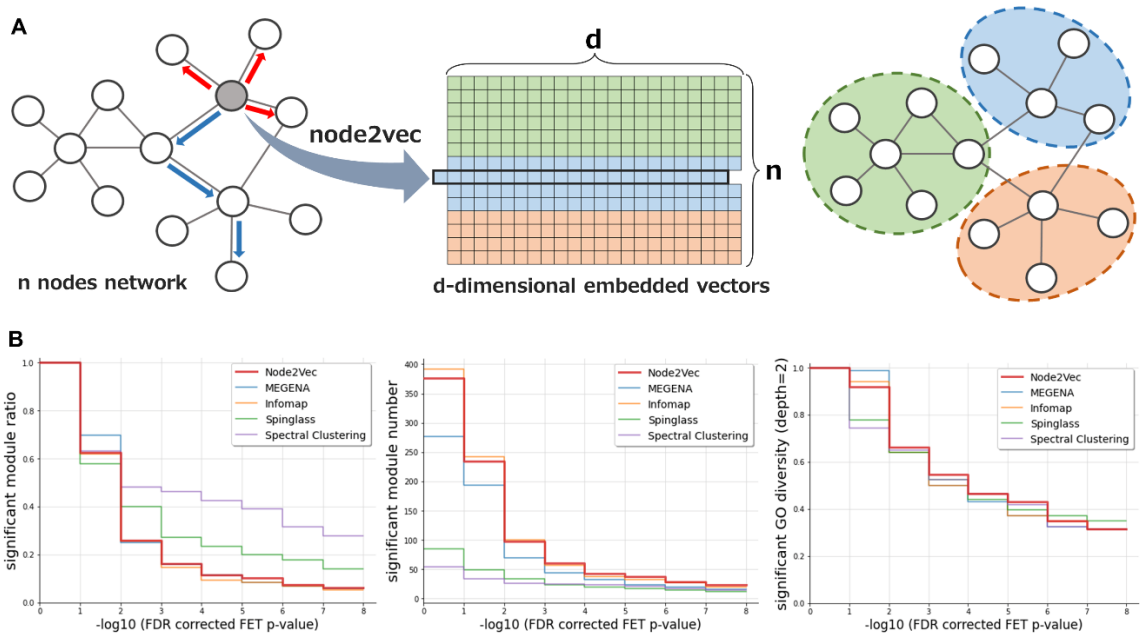

**Fig. 3. Comparison of genetic module detection methods.** Comparison of new module detection methods using node2vec with existing methods in terms of consistency with established biological knowledge. (*A*) Overview of module detection using node2vec. For a GCN consisting of n genes, each node is embedded in a d-dimensional vector to transform an "n × d" matrix. Clustering is applied to the obtained matrix, and the gene sets recognized in the clustering are reflected to the network. (*B*) Comparison of module detection methods in terms of biological plausibility and diversity for the human GBM dataset. Significant module ratio (left), significant module number (middle), and significant GO diversity (right) at significant levels were compared for the five methods including our method using node2vec (bold red line). The *P* values were adjusted for the FDR using the Benjamini–Hochberg method.

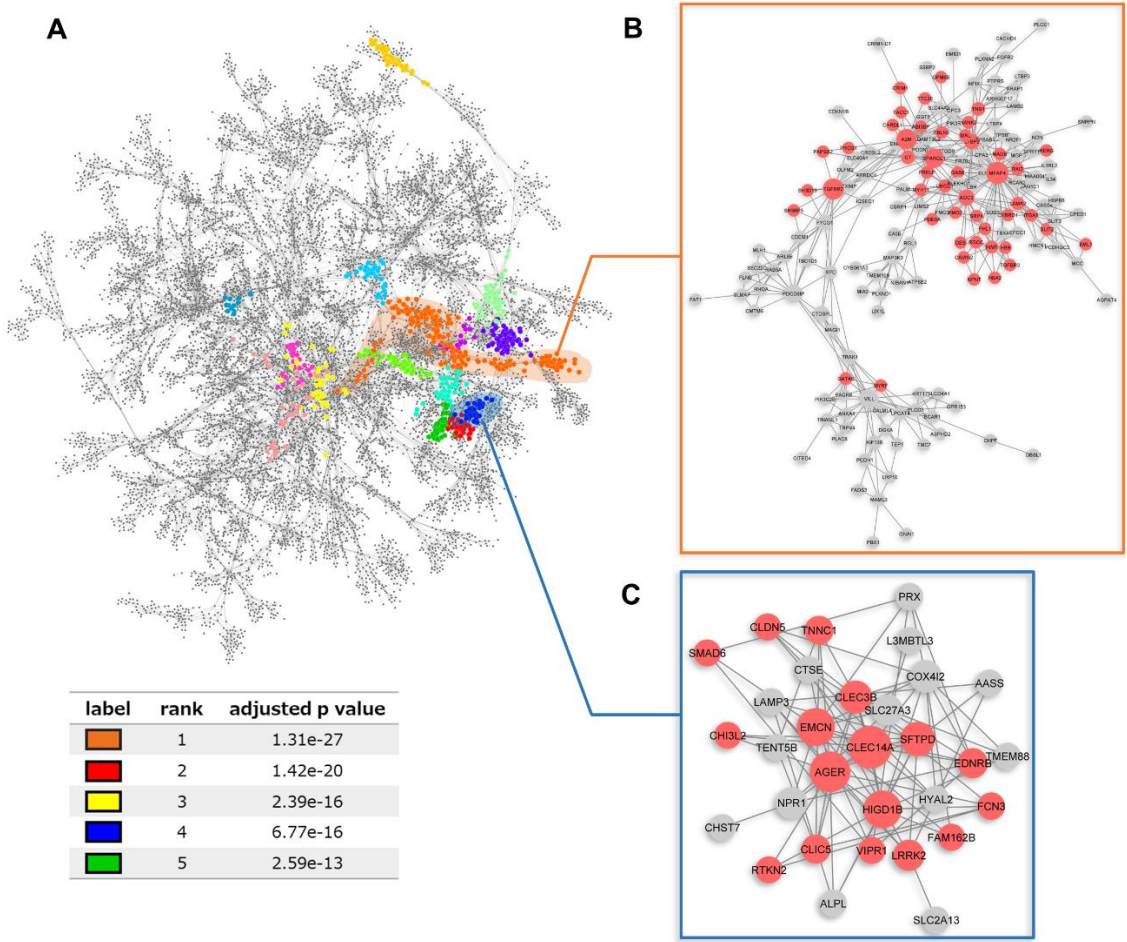

**Fig. 4. Utilization of the predefined gene groups with individual relationships generated by the proposed framework.** (A) The global LUAD GCN constructed with the Correlation-based method. The 15 modules drawn not in gray show a significant overlap with the 742 LUAD DEGs defined by Hou et al. (Hou et al., 2010). Magnified view of the first (B) and fourth (C) modules when each module is ranked according to the FET *P* values of the overlap with DEGs. DEGs are drawn in red, and the size of the node reflects the value of betweenness centrality, which provides a measure of how influential a node in a network.

**Table 1.** Network properties for the TCGA dataset

| disease | methods | average SPD | clustering coefficient | small-worldness |
|---|---|---|---|---|
| GBM | Correlation based | 12.92 | 0.681 | 0.0527 |
| | Tree based | 24.85 | 0.659 | 0.0265 |
| | Linear Regression | 21.80 | 0.574 | 0.0263 |
| LUAD | Correlation based | 14.27 | 0.680 | 0.0476 |
| | Tree based | 30.38 | 0.675 | 0.0222 |
| | Linear Regression | 19.91 | 0.633 | 0.0318 |
| HCC | Correlation based | 12.19 | 0.685 | 0.0562 |
| | Tree based | 18.98 | 0.633 | 0.0334 |
| | Linear Regression | 16.61 | 0.579 | 0.0349 |

**Table 2.** The numbers of TFs in the GBM dataset with significant concentrations of target genes in their proximity

| disease | methods | bootstrap p-value | |
| --- | --- | --- | --- |
| | | < 0.05 | < 0.05 (Bonferroni) |
| GBM | Correlation based | **326** | **209** |
| | Tree based | 246 | 131 |
| | Linear Regression | 184 | 50 |
| LUAD | Correlation based | 256 | 119 |
| | Tree based | 273 | **168** |
| | Linear Regression | **274** | 138 |
| HCC | Correlation based | **181** | **69** |
| | Tree based | 99 | 26 |
| | Linear Regression | 165 | 45 |

**Table 3.** Heatmap visualizing the results of Fig. 3B

| GBM | methods | -log10 (FDR corrected FET p-value) | | | | | | | |
|---|---|---|---|---|---|---|---|---|---|
| | | 1 | 2 | 3 | 4 | 5 | 6 | 7 | 8 |
| ratio | Node2Vec | 0.6250 | 0.2580 | 0.1622 | 0.1144 | 0.1011 | 0.0745 | 0.0612 | 0.0479 |
| | MEGENA | 0.6968 | 0.2491 | 0.1588 | 0.1155 | 0.0830 | 0.0686 | 0.0578 | 0.0433 |
| | Infomap | 0.6189 | 0.2558 | 0.1458 | 0.0946 | 0.0844 | 0.0691 | 0.0512 | 0.0409 |
| | Spinglass | 0.5765 | 0.4000 | 0.2706 | 0.2353 | 0.2000 | 0.1765 | 0.1412 | 0.1412 |
| | Spectral Clustering | 0.6296 | 0.4815 | 0.4630 | 0.4259 | 0.3889 | 0.3148 | 0.2778 | 0.2593 |
| number | Node2Vec | 235 | 97 | 61 | 43 | 38 | 28 | 23 | 18 |
| | MEGENA | 193 | 69 | 44 | 32 | 23 | 19 | 16 | 12 |
| | Infomap | 242 | 100 | 57 | 37 | 33 | 27 | 20 | 16 |
| | Spinglass | 49 | 34 | 23 | 20 | 17 | 15 | 12 | 12 |
| | Spectral Clustering | 34 | 26 | 25 | 23 | 21 | 17 | 15 | 14 |
| diversity | Node2Vec | 0.9186 | 0.6628 | 0.5465 | 0.4651 | 0.4302 | 0.3488 | 0.3140 | 0.2791 |
| | MEGENA | 0.9884 | 0.6395 | 0.5000 | 0.4302 | 0.3721 | 0.3256 | 0.3140 | 0.2558 |
| | Infomap | 0.9419 | 0.6395 | 0.5000 | 0.4651 | 0.3721 | 0.3488 | 0.3140 | 0.2209 |
| | Spinglass | 0.7791 | 0.6395 | 0.5233 | 0.4419 | 0.3953 | 0.3721 | 0.3488 | 0.3372 |
| | Spectral Clustering | 0.7442 | 0.6512 | 0.5233 | 0.4651 | 0.4186 | 0.3256 | 0.3140 | 0.3140 |

**Supplementary Materials**
The file contains supplementary figures and tables in the following order.

1. **Supplementary Figures**
2. **Supplementary Tables**
3. **Supplementary Note**
4. **Supplementary Methods**
5. **References**

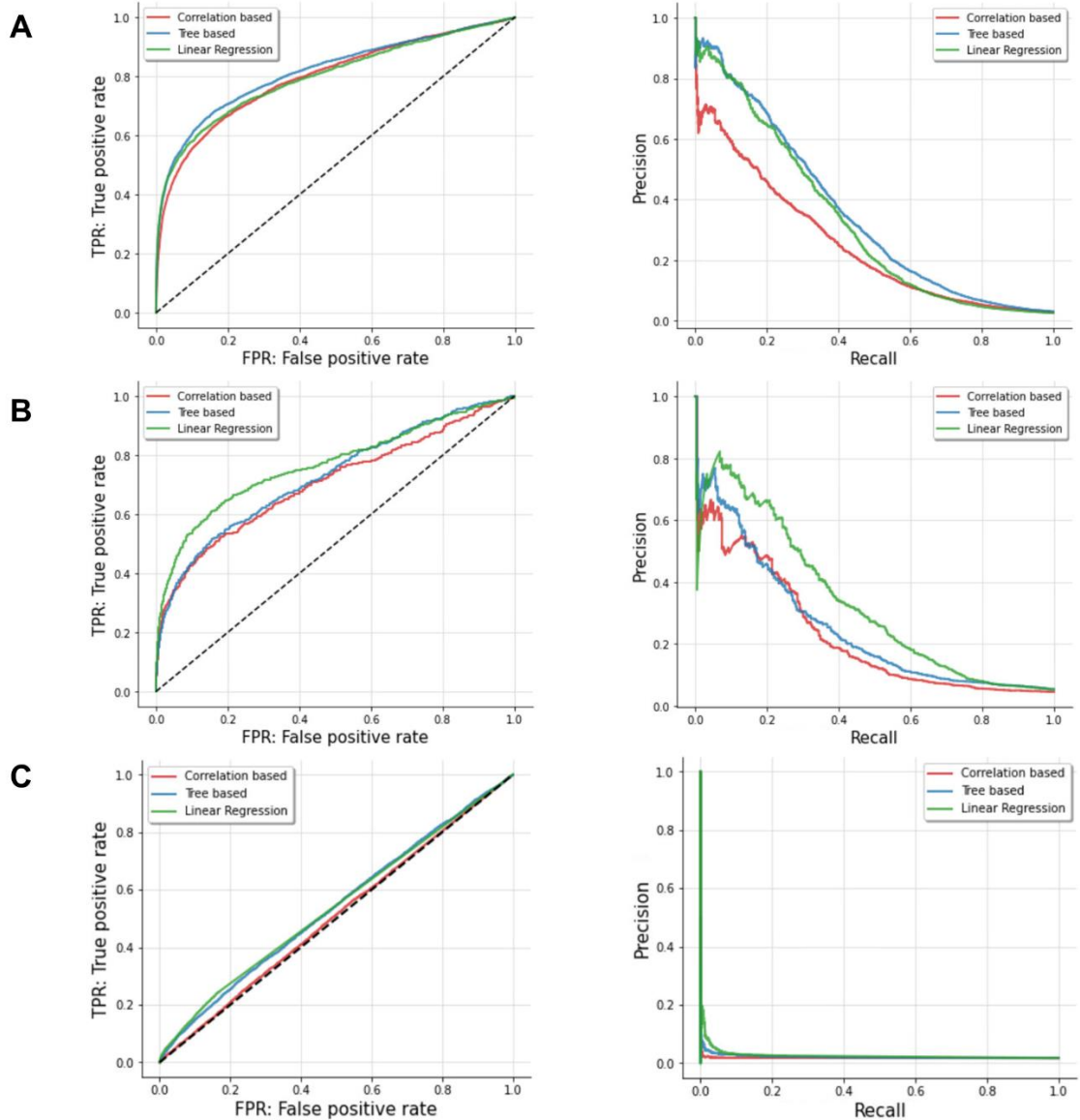

**Supplementary Figure S1. Network inference performance in the DREAM5 dataset**. ROC (left) and PR (right) curves for several methods applied to DREAM5 benchmark networks. These metrics were used to compare the performance of the Correlation-based (red), Tree-based (blue), and Linear Regression (green) in predicting a TF and its target genes' relationships. The number of prediction edges to be analyzed was determined with reference to the DREAM5 challenge. (A) Performance in Network 1 (*in silico*). The top 100,000 edge predictions in its gold standard set were analyzed. (B and C) Performance on Network 3 (*E. coli*, B) and Network 4 (*S. cerevisiae*, C). The top 10,000-edge predictions in its gold standard set were analyzed. Details about AUROC, AUPR, and correctly predicted edge numbers are shown in Supplementary Table S1.

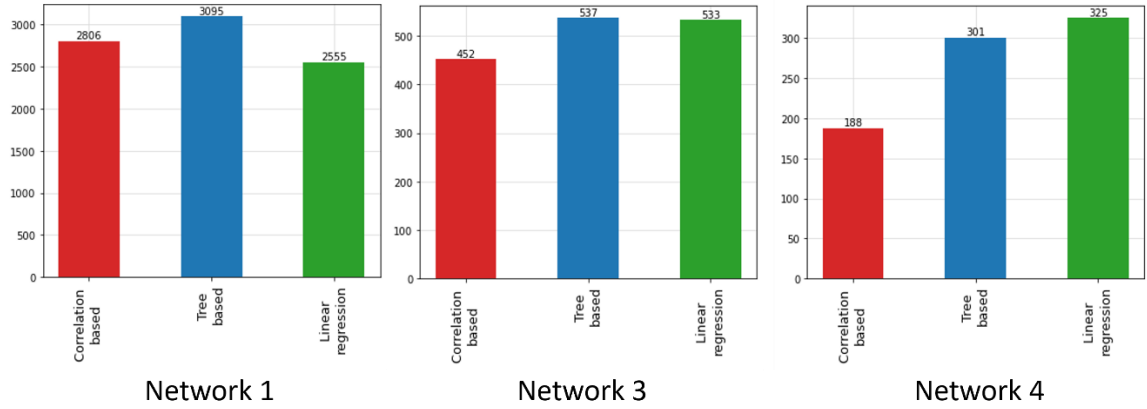

**Supplementary Figure S2. The number of correctly predicted TFs and their target gene pairs in the DREAM5 dataset.** For Network 1, the top 100,000 predicted edges were evaluated while for Network 3 and Network 4, the top 10,000 were evaluated considering the number of total edges. The tree-based method contained more correctly predicted edges than the Correlation-based method for all data sets. In particular, machine learning methods outperformed the Correlation-based one in Network 4.

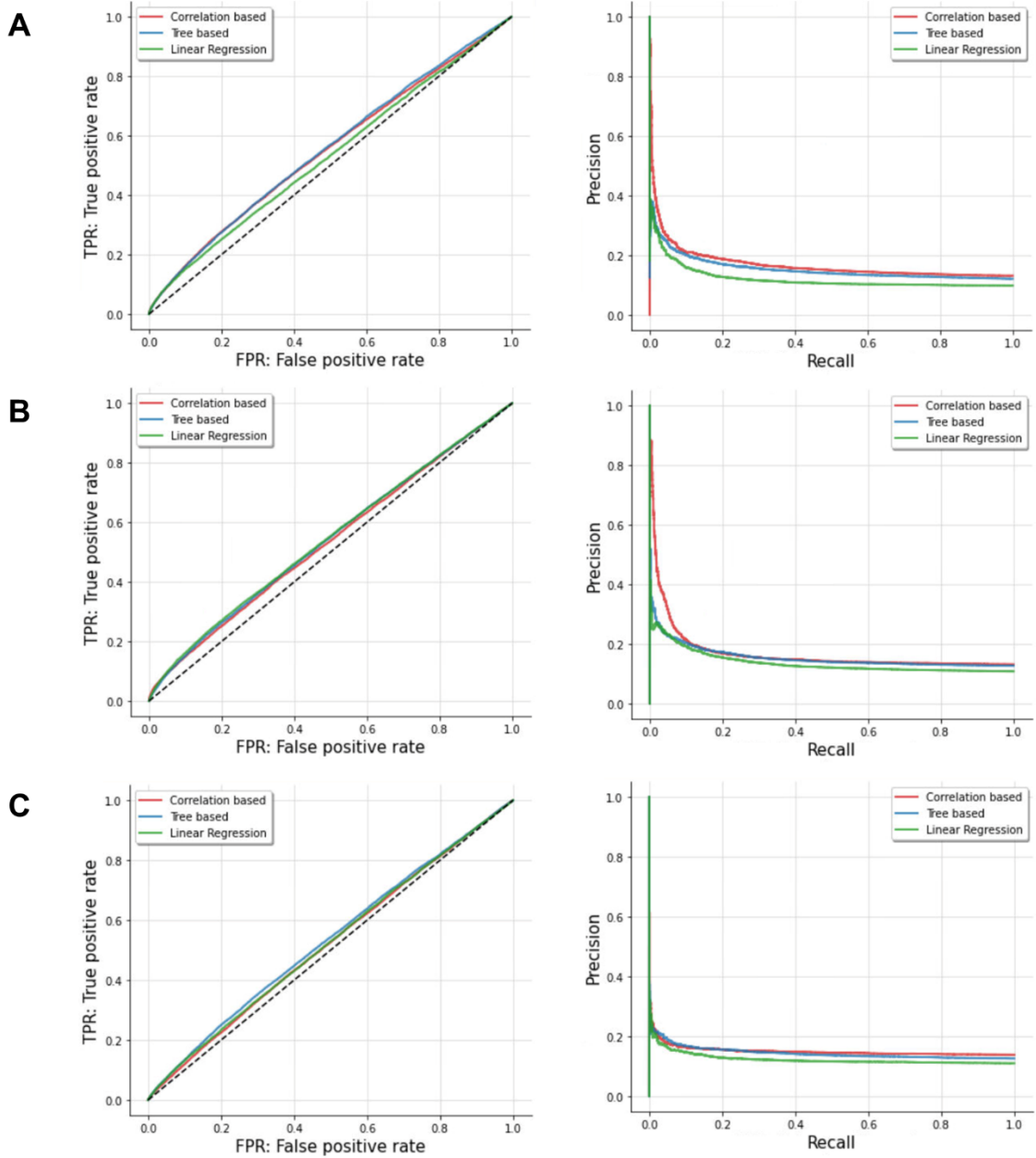

**Supplementary Figure S3. Network inference performance on TCGA dataset.** Prediction performance comparison of Correlation-based (red), Tree-based (blue), and Linear Regression (green) by depicting the ROC (left) and PR (right) curves. As a gold standard, we created a reference based on the ChIP-Atlas dataset. The top 100,000-edge predictions were subjected to evaluation in each dataset from TCGA; (A) glioblastoma (GBM), (B) lung adenocarcinoma (LUAD), and (C) liver hepatocellular carcinoma (HCC). Details about AUROC, AUPR, and correctly predicted edge numbers are listed in Supplementary Table S2.

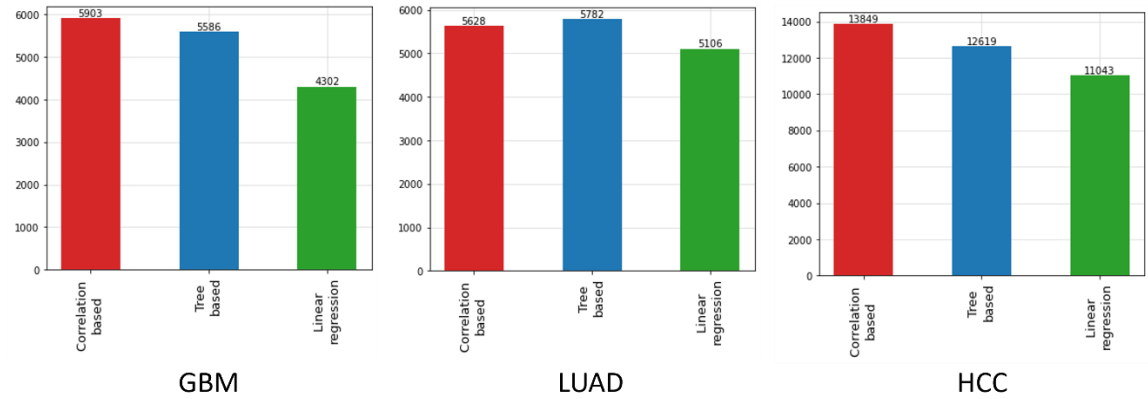

**Supplementary Figure S4. The number of correctly predicted TFs and target gene pairs in the TCGA dataset.** The number of correctly predicted TFs and their target gene pairs in human transcriptome data of GBM, LUAD, and HCC.

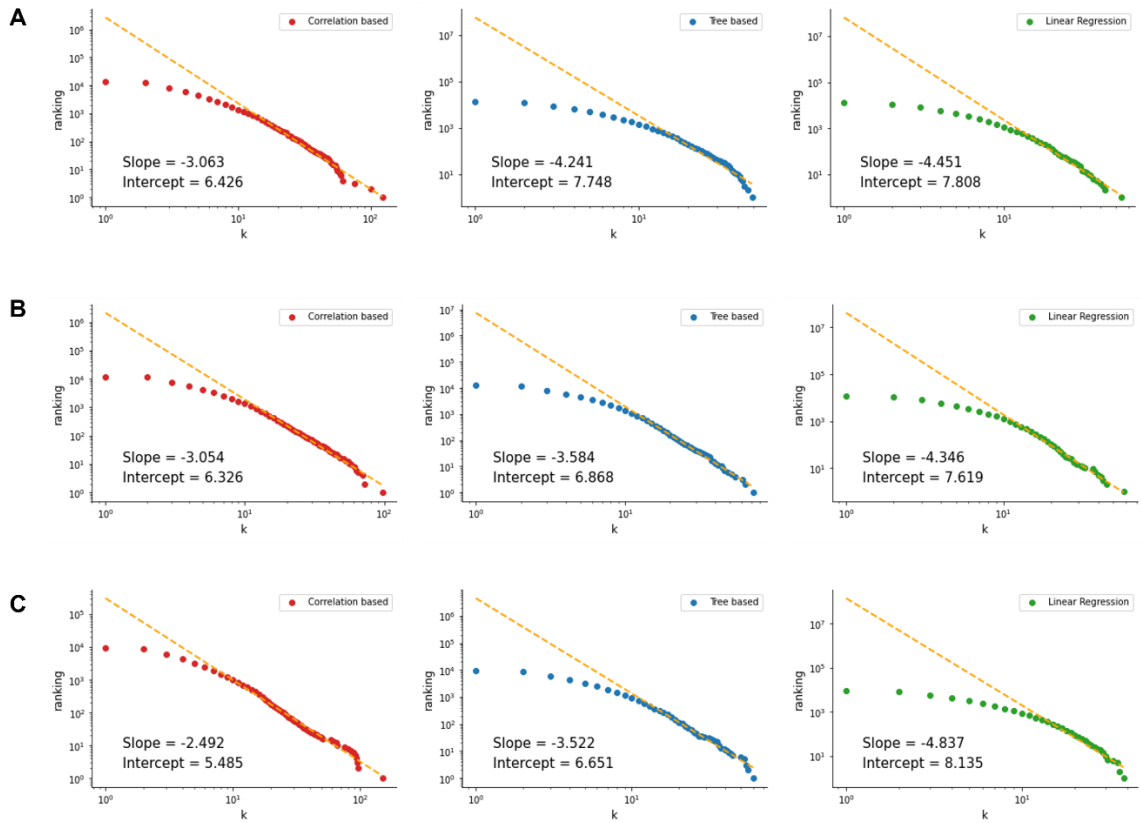

**Supplementary Figure S5. Scale-free property of GCNs derived from the TCGA dataset.** (A–C) Log–log plots of the node degree ranking for GBM, LUAD, and HCC, respectively. The dashed line (orange) is fitted for the region where degree (k) ≥10.

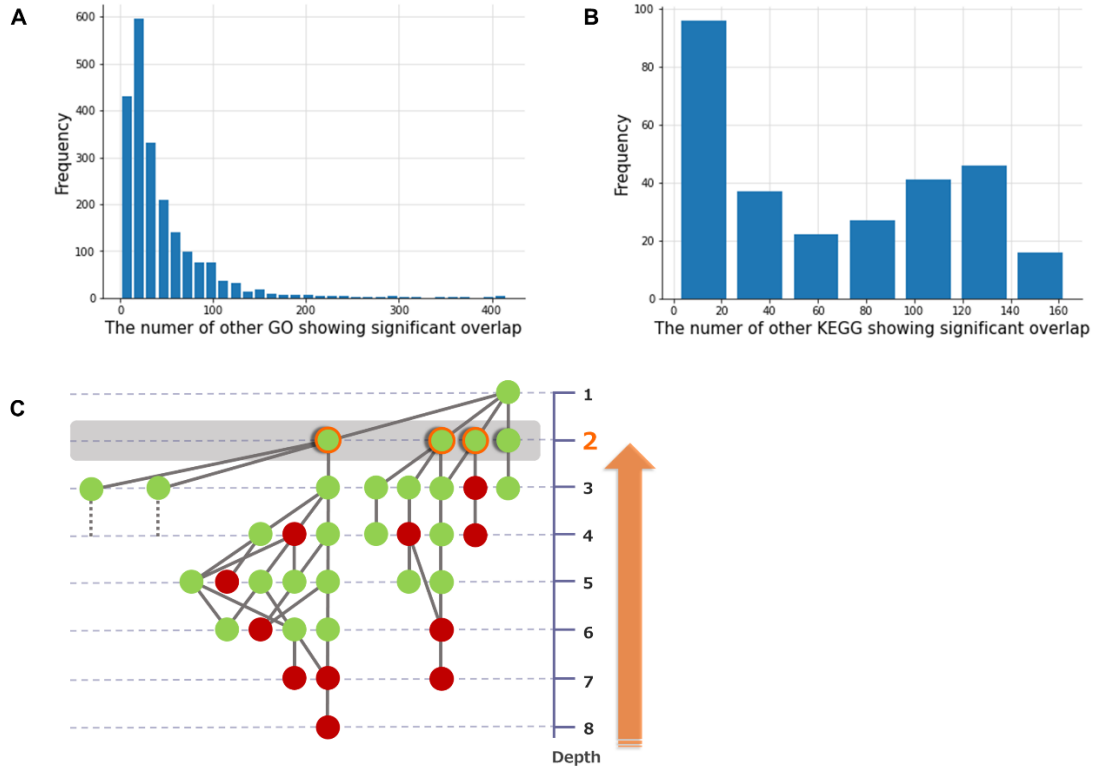

**Supplementary Figure S6. Development of a diversity index for evaluating the genetic module detection method.** The degree of overlap between each gene group and the other registered groups in a database of existing knowledge was evaluated using FET. The histogram shows the number of other groups with significant enrichment (Bonferroni-corrected $P$ value < 0.05). Note that bin size was determined by Scott's rule. (A, B) A curated dataset, 'GO_Biological_Process_2018' and 'KEGG_2019_Human' were obtained from Enrichr. The numbers of registered gene groups were 5,103 and 308, respectively. On average, there were 43.9 and 65.0 duplications in the gene groups registered in GO and KEGG, respectively. (C) GO takes a tree structure but the correspondence is generally evaluated without considering the structure and overlap in the evaluation of modules derived from biological networks. We introduced a diversity index to evaluate the independent biological findings.

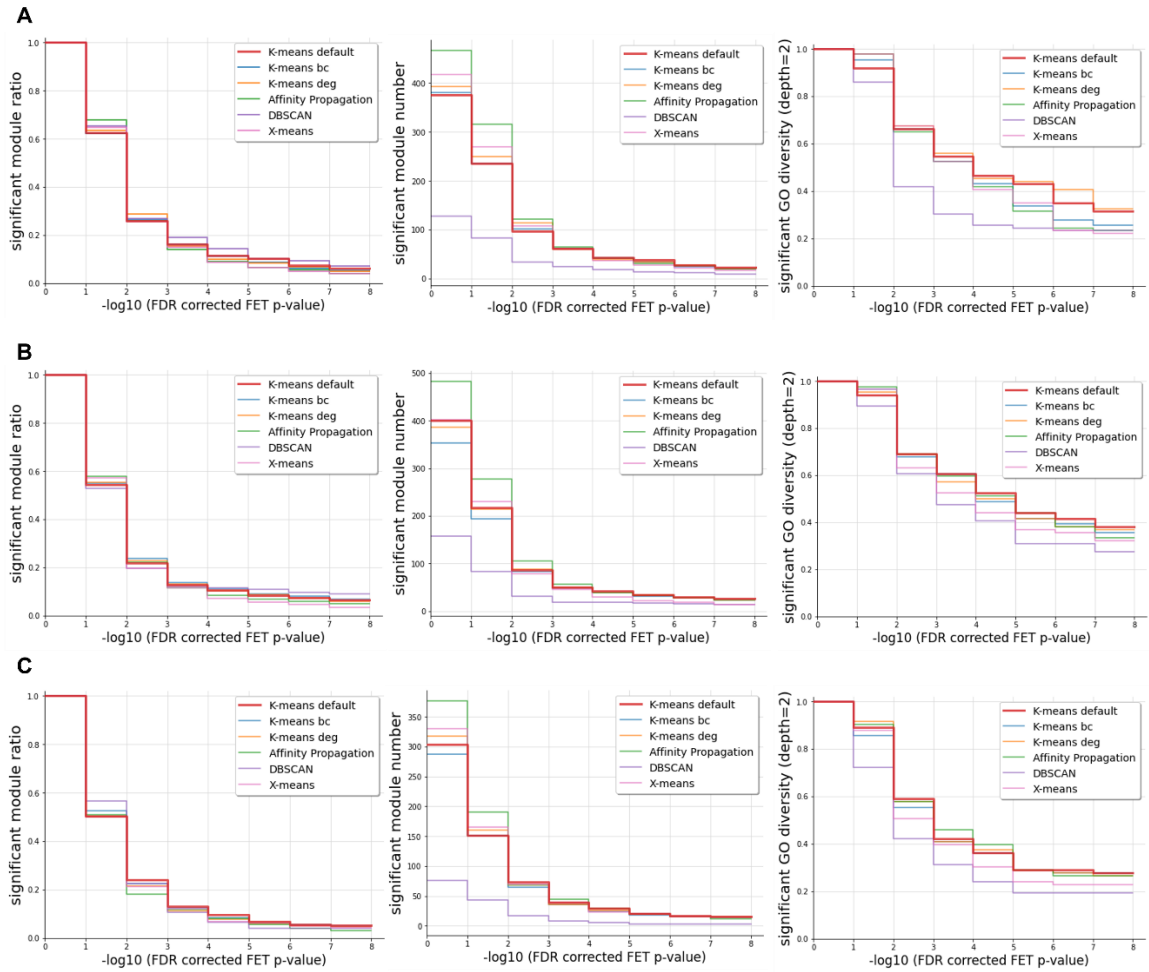

**Supplementary Figure S7. Optimization of clustering methods applying to d-dimensional embedded vectors.** We compared clustering methods in terms of significant module ratios (left), significant module numbers (middle), and significant GO diversities (right) for human GBM (A), LUAD (B), and HCC (C) data.

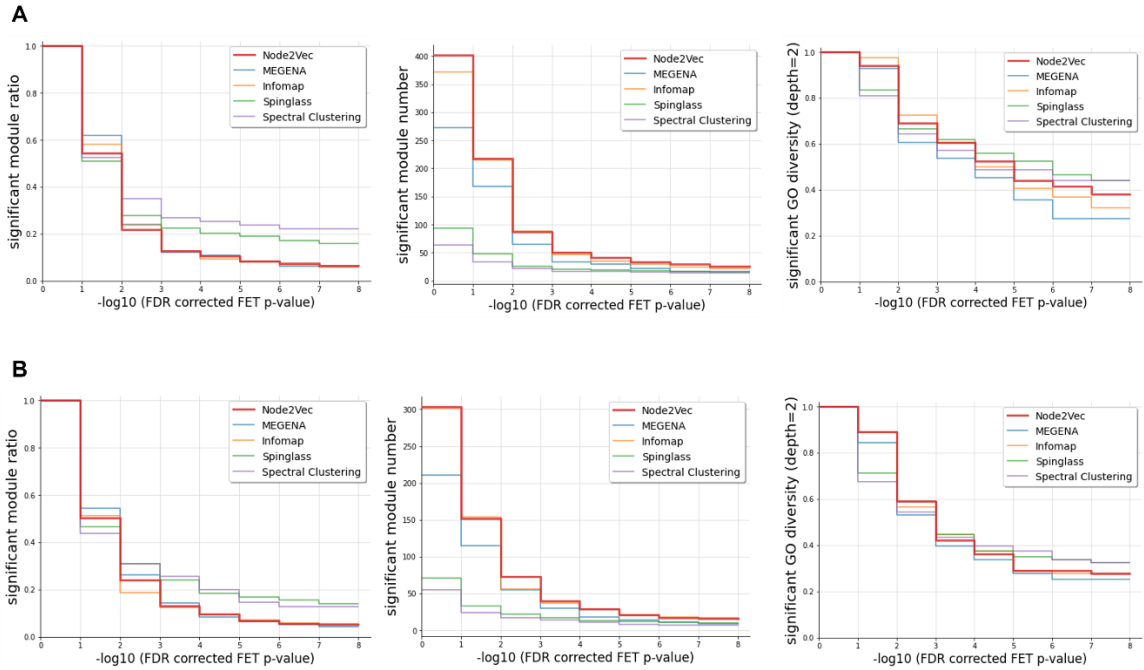

**Supplementary Figure S8. Comparison of module detection methods in terms of biological plausibility and diversity.** Biological plausibility and diversity of modules were evaluated for the human LUAD (A) and HCC (B) datasets.

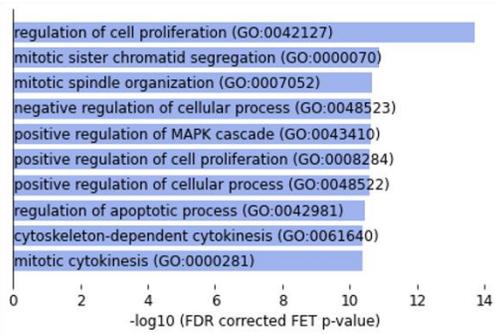 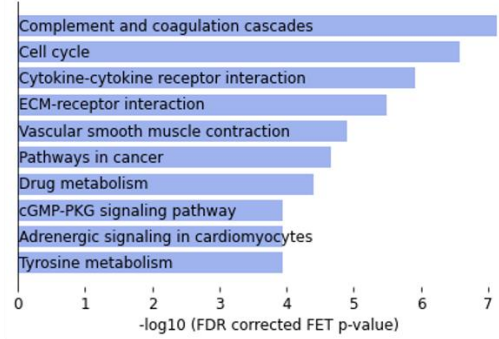

**Supplementary Figure S9. Top 10 FDR-corrected FET *P* values of GO and KEGG pathway analysis for the 742 differentially expressed genes (DEGs).** We regarded 39,000 genes annotated with HGNC as target genes and performed FET for both GO (A) and KEGG (B) pathways.

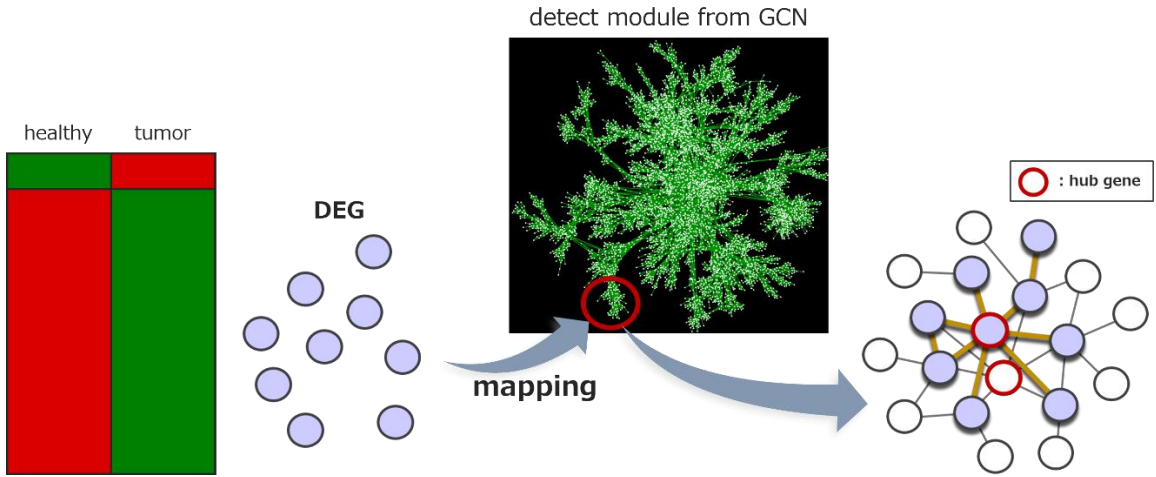

**Supplementary Figure S10. Mapping DEGs on predefined gene groups with individual relationships.** The DEGs derived from a particular disease comprise a group of independent genes. By mapping DEGs to detected modules from GCN, we can add relational information and identify hub genes that are particularly important among the DEGs, or primarily regulate the DEGs.

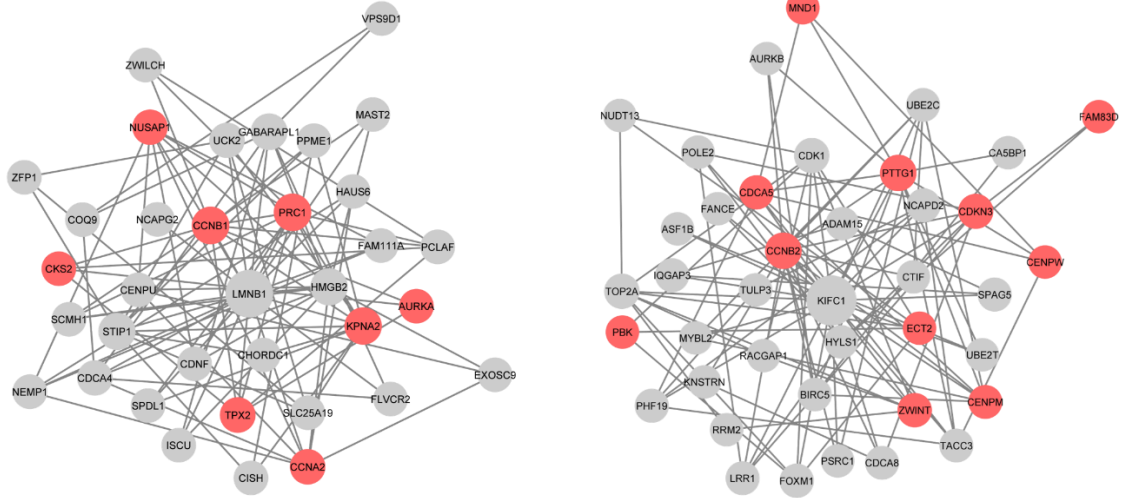
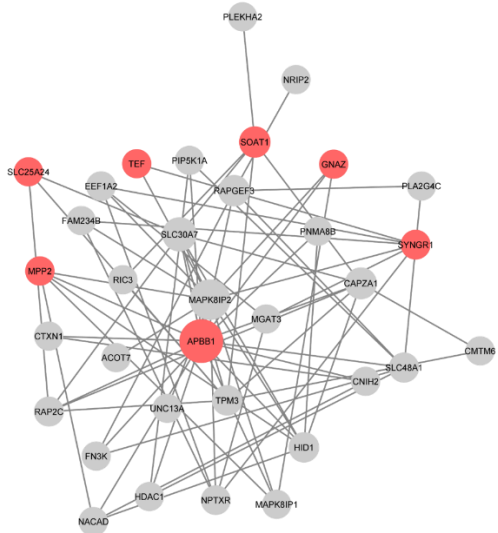

**Supplementary Figure S11. Mapping DEGs on significantly corresponding modules**. The modules that showed significant correspondence with each disease are the indicated DEGs for HCC (A) and GBM (B) data.

**Supplementary Table S1.** Summary of the results of comparing the three inference methods for the benchmark datasets used in the DREAM5 challenge

| Method | Network 1 | | | Network 3 | | | Network 4 | | |
|---|---|---|---|---|---|---|---|---|---|
| | AUROC | AUPR | True label | AUROC | AUPR | True label | AUROC | AUPR | True label |
| Correlation based | 0.796 | 0.249 | 2806 | 0.705 | 0.226 | 452 | 0.507 | 0.0186 | 188 |
| Tree based | **0.816** | **0.352** | **3095** | 0.727 | 0.252 | **537** | 0.538 | 0.0219 | 301 |
| Linear Regression | 0.799 | 0.324 | 2555 | **0.767** | **0.332** | 533 | **0.541** | **0.0263** | 325 |

**Supplementary Table S2.** Summary of the results of comparing the three inference methods for the benchmark datasets used in the TCGA dataset

| Method | GBM | | | LUAD | | | HCC | | |
|---|---|---|---|---|---|---|---|---|---|
| | AUROC | AUPR | True label | AUROC | AUPR | True label | AUROC | AUPR | True label |
| Correlation based | 0.549 | **0.170** | **5903** | 0.534 | **0.169** | 5628 | 0.521 | **0.151** | **13849** |
| Tree based | **0.554** | 0.154 | 5586 | 0.541 | 0.155 | **5782** | **0.534** | 0.145 | 12619 |
| Linear Regression | 0.531 | 0.122 | 4302 | **0.545** | 0.137 | 5106 | 0.524 | 0.124 | 11043 |

**Supplementary Table S3.** The number of TFs whose neighboring genes showed significant overlap with ChIP-Atlas target genes by FET

| disease | methods | proximal secondary | | proximal tertiary | |
|---|---|---|---|---|---|
| | | significan overlap | no overlap | significan overlap | no overlap |
| GBM | Correlatin based | 11 | 253 | 21 | 202 |
| | Tree based | 10 | 298 | 14 | 244 |
| | Linear Regression | 12 | 269 | 12 | 274 |
| LUAD | Correlatin based | 11 | 282 | 19 | 235 |
| | Tree based | 10 | 294 | 15 | 245 |
| | Linear Regression | 11 | 301 | 12 | 298 |
| HCC | Correlatin based | 7 | 219 | 12 | 171 |
| | Tree based | 7 | 238 | 11 | 211 |
| | Linear Regression | 6 | 222 | 8 | 215 |

**Supplementary Table S4**. Heatmap visualizing the results of Supplementary Figure S8.

| LUAD | methods | -log10 (FDR corrected FET p-value) | | | | | | | |
|---|---|---|---|---|---|---|---|---|---|
| | | 1 | 2 | 3 | 4 | 5 | 6 | 7 | 8 |
| ratio | Node2Vec | 0.5436 | 0.2170 | 0.1272 | 0.1047 | 0.0848 | 0.0748 | 0.0648 | 0.0574 |
| | MEGENA | 0.6176 | 0.2390 | 0.1213 | 0.1103 | 0.0809 | 0.0625 | 0.0588 | 0.0551 |
| | Infomap | 0.5795 | 0.2372 | 0.1240 | 0.0943 | 0.0809 | 0.0674 | 0.0593 | 0.0539 |
| | Spinglass | 0.5106 | 0.2766 | 0.2234 | 0.2021 | 0.1915 | 0.1702 | 0.1596 | 0.1489 |
| | Spectral Clustering | 0.5238 | 0.3492 | 0.2698 | 0.2540 | 0.2381 | 0.2222 | 0.2222 | 0.2222 |
| number | Node2Vec | 218 | 87 | 51 | 42 | 34 | 30 | 26 | 23 |
| | MEGENA | 168 | 65 | 33 | 30 | 22 | 17 | 16 | 15 |
| | Infomap | 215 | 88 | 46 | 35 | 30 | 25 | 22 | 20 |
| | Spinglass | 48 | 26 | 21 | 19 | 18 | 16 | 15 | 14 |
| | Spectral Clustering | 33 | 22 | 17 | 16 | 15 | 14 | 14 | 14 |
| diversity | Node2Vec | 0.9405 | 0.6905 | 0.6071 | 0.5238 | 0.4405 | 0.4167 | 0.3810 | 0.3810 |
| | MEGENA | 0.9286 | 0.6071 | 0.5357 | 0.4524 | 0.3571 | 0.2738 | 0.2738 | 0.2738 |
| | Infomap | 0.9762 | 0.7262 | 0.6071 | 0.5000 | 0.4048 | 0.3690 | 0.3214 | 0.3214 |
| | Spinglass | 0.8333 | 0.6667 | 0.6190 | 0.5595 | 0.5238 | 0.4643 | 0.4405 | 0.4286 |
| | Spectral Clustering | 0.8095 | 0.6429 | 0.5714 | 0.4881 | 0.4881 | 0.4405 | 0.4405 | 0.4286 |

| HCC | methods | -log10 (FDR corrected FET p-value) | | | | | | | |
|---|---|---|---|---|---|---|---|---|---|
| | | 1 | 2 | 3 | 4 | 5 | 6 | 7 | 8 |
| ratio | Node2Vec | 0.5017 | 0.2409 | 0.1320 | 0.0957 | 0.0693 | 0.0561 | 0.0528 | 0.0429 |
| | MEGENA | 0.5450 | 0.2607 | 0.1422 | 0.0853 | 0.0664 | 0.0521 | 0.0427 | 0.0427 |
| | Infomap | 0.5116 | 0.1860 | 0.1229 | 0.0930 | 0.0698 | 0.0598 | 0.0565 | 0.0498 |
| | Spinglass | 0.4648 | 0.3099 | 0.2394 | 0.1831 | 0.1690 | 0.1549 | 0.1408 | 0.1268 |
| | Spectral Clustering | 0.4364 | 0.3091 | 0.2545 | 0.2000 | 0.1455 | 0.1273 | 0.1273 | 0.1091 |
| number | Node2Vec | 152 | 73 | 40 | 29 | 21 | 17 | 16 | 13 |
| | MEGENA | 115 | 55 | 30 | 18 | 14 | 11 | 9 | 9 |
| | Infomap | 154 | 56 | 37 | 28 | 21 | 18 | 17 | 15 |
| | Spinglass | 33 | 22 | 17 | 13 | 12 | 11 | 10 | 9 |
| | Spectral Clustering | 24 | 17 | 14 | 11 | 8 | 7 | 7 | 6 |
| diversity | Node2Vec | 0.8916 | 0.5904 | 0.4217 | 0.3614 | 0.2892 | 0.2892 | 0.2771 | 0.2530 |
| | MEGENA | 0.8434 | 0.5301 | 0.3976 | 0.3373 | 0.2771 | 0.2530 | 0.2530 | 0.2530 |
| | Infomap | 0.8916 | 0.5663 | 0.4458 | 0.3735 | 0.2892 | 0.2771 | 0.2771 | 0.2651 |
| | Spinglass | 0.7108 | 0.5904 | 0.4458 | 0.3735 | 0.3494 | 0.3373 | 0.3253 | 0.3012 |
| | Spectral Clustering | 0.6747 | 0.5422 | 0.4337 | 0.3976 | 0.3735 | 0.3373 | 0.3253 | 0.2410 |

**Supplementary Note**
**Comparison of inference methods using the DREAM5 dataset**
Of the three steps from transcriptome data to modules, a planar maximally filtered graph was selected for network construction because it does not require arbitrary hyperparameters (Tumminello *et al.*, 2005). We started by refining the method to infer the relationships between genes. Datasets from the DREAM5 challenge (Marbach *et al.*, 2012), retaining gold standards of transcription factors (TFs) and their target genes, are widely accepted as benchmark datasets for inferring gene relationships. We tested three different types of inference that would cover a large portion of gene–gene relationships. The first one is GENIE3 (Huynh-Thu *et al.*, 2010) as a tree-based machine learning method (Tree-based) because it achieved excellent results in the DREAM5 competition. We considered that other machine learning methods could be useful in relationship inference as well and selected Elastic Net as another approach (Linear Regression) thanks to its robustness against multicollinearity and having a group effect. In addition, we selected an approach that considers the effect of measurement errors in "omics" data by correcting it with the false discovery rate (FDR) as a competitor to the above two types of machine learning (Correlation-based) (Song and Zhang, 2015).

Supplementary Figure S1 shows both the receiver operating characteristic (ROC) curves and the Precision Recall (PR) curves for the three methods, as in the DREAM5 challenge (Supplementary Table S1, Supplementary Note). Both machine learning methods performed better than that of the Correlation-based method in Network 1. For Network 3, Linear Regression showed excellent performance for both AUROC and AUPR measures and there was not much difference between Tree-based and Correlation-based methods, whereas both machine learning methods outperformed the Correlation-based approach in terms of the number of correct edges (Supplementary Figure S2). There was little difference in prediction performance between the three methods in Network 4. The advantage of the Tree-based machine learning method in Network 1 and Network 3, and the difficulty of predicting Network 4 inference using any method were consistent with the DREAM5 challenge results. These results suggest that the two machine learning methods are better than the Correlation-based method in analyses of the DREAM5 datasets.

**Comparison of inference methods with the TCGA dataset**
The datasets used in Supplementary Figure S1 were derived from *in silico* simulation, *Escherichia coli*, and *Saccharomyces cerevisiae* and are far from "real-world" human data. We next compared the performances of the three inference methods using human data from The Cancer Genome Atlas (TCGA) (Weinstein *et al.*, 2013). First, we prepared gold standards for TFs and their target genes using ChIP-Atlas because there is no *de facto* standard (Oki *et al.*, 2018) (Supplementary Methods). The reliability of negative examples is somewhat low because the definition of the target genes for a given TF depends on the properties of the experiments registered in ChIP-Atlas, such as detection sensitivity and the biological background of specimens. However, this gold standard definition is a data-driven approach derived from large-scale data, which allows us to evaluate the inference methods from a perspective different from that of the DREAM5 datasets. ROC and PR curves were drawn for each of the GBM, LUAD, and HCC datasets derived from the TCGA (Supplementary Figure S3, Supplementary Table S2), and the area under the curve was compared. For the AUPR metric, inference using the Correlation-based method corresponded better to ChIP-Atlas for all datasets, and the same was true for the number of correct labels in the top predicted edges (Supplementary Figure S4). However, there was no consistently good method for AUROC, and the prediction accuracy was generally poor. These results imply that widely used evaluations based on prediction accuracy are not always suitable for optimization of the methods to infer the relationships between genes from human data.

**Development of a diversity index for evaluating module detection methods**
In analyses of biological networks, correspondence with predefined gene groups based on domain knowledge is evaluated for comparing module detection algorithms. The ratio of modules that show significant correspondence is often utilized, whereas the results are unstable depending on the number of modules extracted. For this reason, the number of detected modules that show significant correspondence is also used frequently. However, the independence of each gene group is not always guaranteed in the existing databases of gene groups, and the number of significantly correspondent modules could be overestimated. In fact, when we tested for overlap between each of the registered gene groups in several well-known databases, many of the groups exhibited significant overlap with other groups in the same database (Supplementary Figure S6*A*).

To adjust for overlap and cover a large portion of gene–gene relationships, we devised a diversity index that utilizes the structural information of GO definitions. The GO approach uses the tree structure of ontology, and the independence of trees from each other improves as they move to the upper layers (Ashburner *et al.*, 2000). Therefore, by summarizing the GO factors detected in the lower level with those in the upper level, it is possible to define an index of the number of independent biological findings, or diversity, maintaining the power of detecting correspondence with the detected modules (Supplementary Figure S6*B*). In the following analysis, in addition to the conventional metrics, we added diversity as a third indicator to evaluate the performance of module extraction.

## Supplementary Methods
### Data processing
In this study, GCNs were applied to two series of datasets. One comprises the benchmark datasets from the DREAM5 challenge (Marbach *et al.*, 2012), and the other comprises human clinical datasets from TCGA (Weinstein *et al.*, 2013). We used Network 1, Network 3, and Network 4 from the DREAM5 challenge because the gold standards of these methods were available. Human clinical gene expression data of glioblastomas, lung adenocarcinomas, and liver hepatocellular carcinomas were downloaded from TCGA and employed after preprocessing.

### DREAM5
Network 1 comprises *in silico* data simulated using GeneNetWeaver software (Schaffter *et al.*, 2011). Network 3 and Network 4 are real microarray expression data, derived from *E. coli* and *S. cerevisiae*, respectively. Both datasets were first quantile normalized and then z-score normalized across the samples.

### TCGA
Human clinical gene expression data of glioblastomas (GBM), lung adenocarcinomas (LUAD), and liver hepatocellular carcinomas (HCC) were downloaded from TCGA (Weinstein *et al.*, 2013). First, we applied ComBat (Johnson *et al.*, 2007) and corrected the data for batch effects of ethnicity, gender, and site. Then, quantile and z-score normalizations were conducted as well as using the DREAM5 datasets, and finally, log-transformation was performed on the data. Finally, 13,782 genes across 157 samples, 12,628 genes across 499 samples, and 9,843 genes across 145 samples were obtained for GBM, HCC, and LUAD, respectively.

### Definition of human Gold Standard using ChIP-Atlas
For 979 antigens of *Homo sapiens* (hg38), we set ± 5k as "Choose Distance from TSS" and obtained scores about target candidate genes of all experiments registered in ChIP-Atlas (Oki *et al.*, 2018). The average score was calculated for each candidate gene and the values were plotted; 748 transcription factors (TFs) out of 979 were analyzed, relating to more than 100 potential target genes. We normalized the highest and lowest scores, determined the slope of the line connecting the two ends, and defined the point tangent to the same slope as the elbow. Candidates that scored greater than the elbow point were regarded as target genes for each TF. In addition, to define a group of genes that are *not* target genes for TFs, we considered those with scores smaller than the elbow point to be negative examples for the 269 TFs that hold more than 10,000 candidates. Note that the reliability of these negative examples is low because this definition depends on the properties of the experiments registered in ChIP-Atlas such as detection sensitivity and biological background of specimens. These 269 TFs were used to evaluate the binary classification of the performance of predicting the relationship between TFs and their target genes in the relationship inference phase, and all 748 TFs were used to analyze the enrichment of target genes in the proximity of TFs after network construction.

### Machine learning methods preparation
#### Tree-based method
Random Forest is one of the decision tree ensemble methods and can account for nonlinear interactions. We selected GENIE3 (Huynh-Thu *et al.*, 2010) as the tree-based gene relationship estimation method in the R/Bioconductor package. We set default parameters following the GENIE3 example. Specifically, K = sqrt, nCores = 4, nTree = 1000, and seed = 123. Note that the variance in the regression for each gene is equal and normalized because we used the z-score of the input data in this study.

#### Linear Regression method
Elastic Net was implemented using the "scikit-learn" package. After setting $\lambda = 1$ and $\alpha = 0.005$, we performed regression analysis for all genes and obtained the weight of the relationships between each gene. When creating interaction rankings between genes, the sum of the weights for each regression differed, so we corrected the sum to 1 for each regression.

### Complex network properties

A scale-free network shows a power-law degree distribution as follows (Barabási and Albert, 1999):

$$P_{deg}(k) \sim k^\gamma$$

where $\gamma$ is a scale parameter estimated by plotting degree against its rank for each node.

A small world network shows a small average shortest path distance (SPD) and a high clustering coefficient (Humphries and Gurney, 2008). Therefore, we evaluated "small worldness" as follows:

$$S_{ws} = \frac{C_g}{D_g},$$

where $C_g$ is the clustering coefficient and $D_g$ is the average SPD.

In addition to the small worldness property mentioned in Table 1 in the main text, scale-free is another feature of complex networks. After constructing gene co-expression networks (GCNs) for three human gene expression data of GBM, LUAD, and HCC, we compared scale-free properties between three relationship estimation methods: Correlation-based, Tree-based, and Linear Regression. Supplementary Figure S5 shows the log–log plot of node degree ranking. Note that it is reported that few real-world examples follow a power law in the entire range (Newman, 2005), so we assumed that the node degree distribution had a power-law tail in the region of degree (k) ≥10. The estimated exponent $\gamma$ of GCNs using a correlation-based method was close to $2 \leq \gamma \leq 3$, which is the most frequently observed exponent range in complex networks (Albert and Barabási, 2002).

**Evaluation metrics of relationship inference**

We assessed relationship inference performance based on the prediction of the gold standard relationships of TFs and their target genes. Each TF and its target gene predictions were evaluated with ROC and PR curves, which are standard performance metrics in binary classification tasks and are employed in network evaluation (Grover and Leskovec, 2016; Saito and Rehmsmeier, 2015).

**Existing module detection methods to a graph structure**

Methods that detect modules directly in a Top-Down manner using network structure.

**MEGENA**

The module detection method used in MEGENA (Song and Zhang, 2015) is one of the hierarchical-clustering methods and conducted using R. Default values were used for total parameters and modules were retrieved from the lower level so that there was no overlap between them.

**Infomap**

Infomap (Rosvall and Bergstrom, 2008) was implemented using igraph in R with default parameters. Note that it did not work well in the Python environment.

**Spinglass**

Spinglass (Reichardt and Bornholdt, 2006) was implemented using igraph in Python. We set spins = 500 and update_rule = "config." Default values were used for other parameters.

**Spectral clustering**

It is normal to optimize the number of clusters based on the eigenvalue jumping point when it is sorted, but this is difficult to do in large-scale networks such as GCNs. Therefore, we decided on the best split number by updating Newman's modularity value Q in each split. Spectral Clustering (Andrew Y. Ng , Michael I. Jordan, 2001) was implemented using the scikit-learn package.

**Development of node2vec recursive clustering**

Node2vec is a graph-embedding method using the skip-gram algorithm. We applied it to GCNs networks and transformed to vectors. The parameter settings used in this study were decided based on the case applying to BlogCatalog (Liu, 2009). Specifically, we set d = 128, r = 20, l = 100, k = 15, p = 1, and q = 1. The recursive clustering algorithm of the network based on

Newman's modularity (Q) was inspired by Song et al. (Song and Zhang, 2015) and is applied as follows:
1. Transform the GCN to d-dimensional feature vectors about n genes with node2vec.
2. Repeat k-means clustering with increasing the k-split number one by one and calculate Newman's modularity Q in each split until no better value is returned for 10 consecutive times. The k-split maximizing Q is the best split number and allows this k-means clustering.
3. For the clusters defined above, we perform hierarchical clustering recursively. In this step, the minimum cluster size threshold and the compactness degree are considered to determine whether the next k-means split is allowed. Specifically, after splitting with optimal k-split by maximizing modularity Q, check for each new cluster as to whether it meets the minimum cluster size threshold and calculate the compactness degree calculated as follows:
$$v_l = \frac{SPD \in V_l}{\log(|V_l|)}.$$
The split is rejected if the percentage of new clusters that meet the minimum cluster size is less than 80% or if all new clusters are not compact compared with the parent cluster.

Note that various unsupervised clustering methods could be applied to the network vectorized with "node2vec."

**Analysis of differentially expressed genes (DEGs) by predefined gene groups**
Predefined gene groups with relationships can complement the relationship between individual genes in a gene group such as DEGs and support a deeper understanding of the molecular machinery behind the particular gene group. For GCNs constructed with human GBM, LUAD, and HCC data from TCGA, we evaluated the usefulness of detected modules to similar physiological DEGs. To investigate the ability to extrapolate, we obtained DEG information from independent datasets of the data used in TCGA. Specifically, we retrieved 742 DEGs of LUAD from the report of Hou et al. (Hou *et al.*, 2010), 89 DEGs of HCC from the report by Song et al. (Song *et al.*, 2019), and 263 DEGs of GBM from Chen et al. (Chen *et al.*, 2020). We applied Fisher's Exact Test (FET) to evaluate the overlap between the DEGs and the detected modules. Note that in this analysis, the total number of target genes for FET was equal to the number of genes in the GCN.

**Evaluation metrics of relationship inference**
Precision and recall are expressed as follows:

$$\text{recall}(k) = \frac{TP(k)}{P},$$
$$\text{precision}(k) = \frac{TP(k)}{TP(k) + FP(k)},$$

where TP(k) and FP(k) are the true positive and false positive numbers in the top k predictions of the ranked interaction list, respectively. P is the number of positives in the gold standard interaction.

**Applying unsupervised clustering methods to a vectorized matrix**
In addition to the method described above, we prepared several clustering methods that do not require prespecifying the number of clusters, k. Specifically, we applied Affinity Propagation (Frey and Dueck, 2007), DBSCAN (Ester M, Kriegel H-P, Sander J, 1996), and X-means (Dau Pelleg, 2000).
Detailed information about these three clustering methods was implemented using the scikit-learn packages, and X-means were derived from pyclustering.
**Affinity Propagation**
Default values were used for all parameters except for the random seed, which was set to 123.
**DBSCAN**
We set eps = 0.3 and min_samples = 10. Default values were used for the other parameters.
**X-means**

After optimizing the initial centers with kmeans_plusplus_initializer class, we conducted X-means clustering with kmax set to 1000 and the rest of the parameters were set to the default values.